\documentclass{ws-procs975x65}

\newcommand{\beqs}{\begin{equation*}}
\def\beq{\begin{equation}}

\newcommand{\eeqs}{\end{equation*}}
\def\eeq{\end{equation}}

\newcommand{\beqas}{\begin{eqnarray*}}
\newcommand{\beqa}{\begin{eqnarray}}

\newcommand{\eeqas}{\end{eqnarray*}}
\newcommand{\eeqa}{\end{eqnarray}}

\newcommand{\eq}[2]{\begin{equation} #1 \label{#2} \end{equation}}

\newcommand{\eps}{\varepsilon}

\newcommand{\ga}{\gamma}
\newcommand{\de}{\delta}
\newcommand{\om}{\omega}

\newcommand{\la}{\lambda}

\newcommand{\Om}{\Omega}

\newcommand{\blist}{\begin{itemize}}

\newcommand{\elist}{\end{itemize}}

\providecommand{\href}[2]{#2}

\DeclareFontFamily{OT1}{rsfs}{}
\DeclareFontShape{OT1}{rsfs}{m}{n}{ <-7> rsfs5 <7-10> rsfs7 <10->rsfs10}{} 
\DeclareMathAlphabet{\mycal}{OT1}{rsfs}{m}{n}

\newcommand{\chib}{\overline{\chi}}

\usepackage{slashed}

\def\cM{{\cal M}}

\def\extd{{\rm d}}

\newcommand{\dirac}[2]{\{ #1 , #2 \}^\ast}

\def\atbdry{\Big|_{\partial \cM}}
\def\atbdry0{\Big|_{\partial \cM_0}}
\def\atbdry1{\Big|_{\partial \cM_1}}

\begin{document}

\markboth{R. Meyer}{Quantizing Two-Dimensional Dilaton Gravity with Fermions: The Vienna Way}

\wstoc{Quantizing Two-Dimensional Dilaton Gravity with Fermions: The Vienna Way}{R. Meyer}

\title{Quantizing Two-Dimensional Dilaton Gravity with Fermions: \\The Vienna Way}

\author{Ren\'e Meyer\footnote{Email:~\email{meyer@mppmu.mpg.de}}}

\address{
Max-Planck Institute for Physics,
Werner-Heisenberg Institute,\\ 
F\"ohringer Ring 6, D-80805 M\"unchen, Germany\\ and\\
Institute for Theoretical Physics, 
University of Leipzig, \\
Augustusplatz 10-11, 
D-04103 Leipzig, Germany\\
}


\begin{abstract}
I review recent work on nonperturbative path integral quantization of two-dimensional dilaton gravity coupled to Dirac fermions, employing the "Vienna school`` approach.
\end{abstract}

\bodymatter

Despite much progress in our knowledge of quantum gravity\cite{Carlip:2001wq} during the last  decades, a fully satisfactory quantization of the simplest nontopological\footnote{In the sense that it possesses physical locally propagating degrees of freedom, \textit{i.e.} gravitons.} gravity theory, namely general relativity in four dimensions, is still missing. The reasons  are two-fold: On one hand, standard techniques from perturbative quantum field theory do not apply  to arbitrary high energies to perturbatively nonrenormalizable general relativity. On the other hand, its highly nonlinear dynamics makes general relativity hard to approach with nonperturbative methods. Adding the lack of observational data for quantized gravitational effects, it is hard to compare the suitability of different approaches and methods to quantize gravity. In such a context it may be useful to consider less complicated situations, where even conservative methods like standard quantum field theory can be applied to gravity. 

Such a situation is given in lower dimensions. In two dimensions, however, pure Einstein-Hilbert gravity is topological, the action being proportional to the Euler number.  One way of constructing a two-dimensional gravity theory with sensible dynamics is to add an additional scalar field, henceforth called the dilaton $X$, to Einstein-Hilbert gravity, and possible matter. This leads to the vast subject of two-dimensional dilaton gravity\cite{Grumiller:2002nm}. Such models arise from spherical reduction of general relativity, from string theory as well as as toy models for intrinsic two-dimensional gravity. 

Of the many interesting features of these theories, 
I focus on the application of the nonperturbative path integral quantization method, developed by the ''Vienna school'' around Wolfgang Kummer \cite{Kummer:1996hy}
, to dilaton gravity coupled to Dirac fermions\cite{Meyer:2005fz,Grumiller:2006ja,meyer06}. This method relies on several crucial points: First, using the spin connection $\om=\om_\mu\extd x^\mu$ and dyad 1-forms $e^a=e^a_\mu \extd x^\mu$ built from the inverse Zweibeine $e_\mu^a$, the action for Generalized Dilaton Theories\footnote{$U,V$ parametrize different models (\textit{cf.} tab.~1 in \cite{Grumiller:2006ja}). Notation and conventions are chosen according to {\cite{Grumiller:2006ja}}.},
\begin{equation}
\label{eq:GDT}
S^{(2)}=-\frac12 \int_{\mathcal{M}_2} \extd^{2}x\, \sqrt{-g}\; \left[ X R + U(X)\; (\nabla X)^{2} - 2V(X)\; \right] + S^{(m)}\,,
\end{equation}
is reformulated as a First Order Gravity  action\footnote{$X_a$ are Lagrange multipliers for the torsion $T^a=\extd e^a + \eps^a{}_b \om\wedge e^b$ ($\om^a{}_b = \eps^a{}_b\om$ in two dimensions.), $\eps_{ab}=-\eps_{ba}$, $\eps_{01}=+1$ and $\epsilon=\sqrt{-g}\extd^2 x$ denotes the volume 2-form. The Ricci scalar is $R=-2\ast \extd \om$, where $\ast$ is the Hodge star operator.}
\beqa\label{eq:FOG}
S^{\rm (1)} & = & \int_{\mathcal{M}_2} \left[X_a T^a+X\extd \om+\epsilon\left( U(X)\frac{X^aX_a}{2} + V(X) \right)\right] + S^{(m)}\,.
\eeqa
If the matter in $S^{(m)}$ does not couple to the auxiliary fields $\om$ and $X_a$, these fields can be integrated out \textit{s.t.} \eqref{eq:FOG} and \eqref{eq:GDT} are equivalent both on the classical and quantum level\footnote{\textit{Cf. e.g.} the first paper in\cite{Kummer:1996hy}~.}. This is the case for scalar fields as well as intrinsic two-dimensional Dirac fermions ($a\overleftrightarrow{\extd}b = a(\extd b)-(\extd a) b$)
\eq{S^{(m)} = \int_{{\cal M}_2} \Big[\frac{i}{2} F(X)\;(*e^a) \wedge (\chib \ga_a \overleftrightarrow{\extd}\chi) - \epsilon H(X)\left( m \chib\chi + \la (\chib\chi)^2 \right)\Big] \,,}{eq:fermions}
 but not for spherically reduced four-dimensional fermions\cite{Balasin:2004gf}.  The functions $F,H$ are generic dilaton couplings, and the most general self-interaction for fermions in two dimensions contains at most a quartic term. A second crucial point is the use of light cone gauge for the local \textit{Lorentz} frame, \textit{e.g.} $X^\pm = (X^0 \pm X^1)/\sqrt{2}$.

The path integral quantization of \eqref{eq:FOG} and \eqref{eq:fermions} then consists of four steps: 
\textbf{1. Constraint Analysis
} The system possesses two diffeomorphisms and the local $\mathrm{SO}(1,1)$ symmetry. They are  generated on-shell by three first class constraints $G_i$, which form a nonlinear Lie algebra\footnote{$\dirac{f}{g}$ is the Dirac bracket, taking care of the usual second class constraints in the presence of fermions.} $\dirac{G_i(x)}{G_j(y)} = f_{ij}{}^k(x) G_k \de(x-y)$ with field-dependent structure functions $f_{ij}{}^k(x)$. \eqref{eq:FOG} and \eqref{eq:fermions} thus behaves like a nonlinear Yang-Mills theory rather than a gravity theory, in which the constraint algebra would typically close with derivatives of delta functions\footnote{The classical Virasoro algebra, \textit{i.e.} the one without central charge, is recovered by field-dependent linear combinations of the $G_i$.}.
\textbf{2. BVF Formalism.}\cite{Fradkin:1975cq}
 Accounting for the three gauge symmetries, one introduces three (anti)ghosts $(c_i, p_j^c)$, $i,j=1,2,3$. The BRST charge takes the form as for a Yang-Mills theory, $\Om = c^i G_i + \frac{1}{2}c^i c^j {f_{ij}}^k(x) p_k^c$. With the gauge fixing fermion $\Psi=p_2^c$, axial (or \textit{Eddington-Finkelstein}) gauge  $(\om_0,e_0^-,e_0^+)=(0,1,0)$ is  reached.
~\textbf{3. Nonperturbative Path Integral Quantization of the Geometric Sector} 
The phase space path integral is then evaluated follows: 1.~ Integration over the (anti)ghosts  yields the Faddeev-Popov determinant, solely depending on $(X,X^\pm)$. 2.~In the chosen gauge, the action depends linearly on $(\om_1,e_1^\mp)$ \textit{s.t.} this integration can be carried out directly, yielding delta functionals in the path integral which contain the classical equations of motion for the $(X,X^\pm)$. These equations still include (up to that point still off-shell) fermion terms. 3.~Integrating out $(X,X^\pm)$ then sets these fields to their on-shell values, where the fermion terms are viewed as off-shell external sources. During this step, the Faddeev-Popov determinant cancels, \textit{i.e.}, as typical for axial gauges, the ghosts decouple. Because the equations of motion for $(X,X^\pm)$ are solved using classical Green functions, the asymptotic geometry has to be fixed  and thus an asymptotic Fock space can be constructed. The  quantum fields  $(X,X^\pm)$ fulfill the classical equations of motion before integrating out the fermions because no physical locally propagating degrees of freedom that could yield quantum corrections are present in the geometric sector. \textbf{4. Matter Perturbation Theory} The effective action obtained so far is nonlocal in space but local in time, and nonpolynomial in the fermions. Carrying out the path integration over the fermions  perturbatively generically yields effective nonlocal $2n$-point vertices.

\textbf{Some Results and Outlook} Reminiscent of bosonization in two flat dimensions\cite{Coleman:1974bu}
, two of the three four-fermi vertices\cite{Grumiller:2006ja} coincide with the two effective four-boson vertices found in a similar analysis for scalar fields\cite{Grumiller:2002dm}, while the third one vanishes for on-shell external momenta. However, the asymptotic modes for bosons and fermions differ. In order to investigate bosonization in quantum dilaton gravity, one thus has to compare observables, \textit{e.g.} the four-particle S-matrices of the fermionic and the known bosonic case\cite{Fischer:2001vz} or the specific heat of the Witten black hole (or CGHS model)\cite{Grumiller:2003mc}.
From the scalar case\cite{Fischer:2001vz} one also expects unitarity, \textit{i.e.} no information loss, and CPT invariance of the S-matrix. 

The whole quantization procedure is background independent and only uses standard quantum field theory methods. In order to recover the correct semiclassical limit one also has to sum over degenerate metrics in the path integral. Another interesting application would be to reconstruct black holes as macroscopic bound states of quantum dilaton gravity in a Bethe-Salpeter\cite{Salpeter:1951sz} like manner.
%

\begin{thebibliography}{10}

\bibitem{Carlip:2001wq}
S.~Carlip, {\em Rept. Prog. Phys.} {\bf 64}, p. 885 (2001).

\bibitem{Grumiller:2002nm} Reviews: 
D.~Grumiller, W.~Kummer and D.~V. Vassilevich, {\em Phys. Rept.} {\bf 369}, 327
  (2002);
%
D.~Grumiller and R.~Meyer (2006), \textit{hep-th/0604049}.

\bibitem{Kummer:1996hy}
W.~Kummer, H.~Liebl and D.~V. Vassilevich, {\em Nucl. Phys.} {\bf B493}, 491
  (1997), 
%
 {\bf B513}, 723
  (1998) and
%
 {\bf B544}, 403
  (1999); 
%
D.~Grumiller, 
PhD thesis, {T}echnische {U}niversit{\"a}t {W}ien (2001), \textit{gr-qc/0105078}; 
%
L.~Bergamin, D.~Grumiller and W.~Kummer, {\em JHEP} {\bf 05}, p. 060 (2004); 
%
L.~Bergamin (2004), \textit{hep-th/0408229}; 
%
L.~Bergamin, D.~Grumiller, W.~Kummer and D.~V. Vassilevich, {\em Class. Quant.
  Grav.} {\bf 22}, 1361 (2005).

\bibitem{Meyer:2005fz}
R.~Meyer (2005), \textit{hep-th/0512267}.

\bibitem{Grumiller:2006ja}
D.~Grumiller and R.~Meyer, {\em Class. Quant. Grav.} {\bf 23}, 6435 (2006).

\bibitem{meyer06}
R.~Meyer, 
   Master's thesis, {U}niversit{\"a}t {L}eipzig (2006), \textit{gr-qc/0607062}.

\bibitem{Balasin:2004gf}
H.~Balasin, C.~G. Boehmer and D.~Grumiller, {\em Gen. Rel. Grav.} {\bf 37},
  1435 (2005).

\bibitem{Fradkin:1975cq}
E.~S. Fradkin and G.~A. Vilkovisky, {\em Phys. Lett.} {\bf B55}, p. 224 (1975); 
%
I.~A. Batalin and G.~A. Vilkovisky, {\em Phys. Lett.} {\bf B69}, 309 (1977); 
%
E.~S. Fradkin and T.~E. Fradkina, {\em Phys. Lett.} {\bf B72}, p. 343 (1978).


\bibitem{Grumiller:2002dm}
D.~Grumiller, W.~Kummer and D.~V. Vassilevich, {\em European Phys. J.} {\bf
  C30}, 135 (2003).

\bibitem{Coleman:1974bu}
S.~R. Coleman, {\em Phys. Rev.} {\bf D11}, p. 2088 (1975); 
%
S.~R. Coleman, R.~Jackiw and L.~Susskind, {\em Ann. Phys.} {\bf 93}, p. 267
  (1975).

\bibitem{Fischer:2001vz}
P.~Fischer, D.~Grumiller, W.~Kummer and D.~V. Vassilevich, {\em Phys. Lett.}
  {\bf B521}, 357 (2001), Erratum ibid. {\bf B532} (2002) 373.

\bibitem{Grumiller:2003mc}
D.~Grumiller, W.~Kummer and D.~V. Vassilevich, {\em JHEP} {\bf 07}, p. 009
  (2003).

\bibitem{Salpeter:1951sz}
E.~E.~Salpeter and H.~A.~Bethe, {\em Phys. Rev.} {\bf 84}, p. 1232 (1951)

\end{thebibliography}

\vfill
\pagebreak

\end{document}